\documentclass[useAMS,usenatbib]{mn2e}

\usepackage {graphicx}

\title[Oxygen, $\alpha$-element and iron abundance distributions 
in the inner part of the Galactic thin disc]
{Oxygen, $\alpha$-element and iron abundance distributions in 
the inner part of the Galactic thin disc
\thanks{Based on observations obtained at the Canada-France-Hawaii 
Telescope (CFHT), which is operated by the National Research 
Council of Canada, the Institut National des Sciences de l'Univers 
of the Centre National de la Recherche Scientifique of France, 
and the University of Hawaii.}
\thanks{Based on observations made with ESO Telescopes at the la 
Silla and Paranal Observatories under programmes 089.D0489 and 
60A-9700.}}

\author[R.P. Martin et al.]
{R.P. Martin$^{1}$\thanks{E-mail: rpm33@hawaii.edu},
S.M. Andrievsky$^{2,3}$,
V.V. Kovtyukh$^{2}$,  
S.A. Korotin$^{2}$,
I.A. Yegorova$^{4}$, 
\newauthor
and Ivo Saviane$^{4}$,
\\
$^1$Department of Physics and Astronomy, University of Hawai'i at Hilo, Hilo, 
HI, 96720, USA\\
$^2$Department of Astronomy and Astronomical Observatory, Odessa
National University\\ and Isaac Newton Institute of Chile, 
Odessa Branch, Shevchenko Park, 65014 Odessa, Ukraine\\
$^3$GEPI, Observatoire de Paris-Meudon, CNRS, Universite Paris Diderot, 92125
Meudon Cedex, France\\
$^4$European Southern Observatory, Alonso de Cordova 3107, Santiago,  Chile\\
}

\begin{document} 

\date{Accepted, 12 March 2015. Received, 12 March 2015; in original form, 18 January 2015 }

\pagerange{\pageref{firstpage}--\pageref{lastpage}} \pubyear{2015}

\maketitle

\label{firstpage}

\begin{abstract}
We derived elemental abundances in 27 Cepheids, the great majority 
situated within a zone of Galactocentric distances ranging from 5 to 
7 kpc. One star of our sample, SU Sct, has a Galactocentric distance 
of about 3 kpc, and thus falls in a poorly investigated region of the 
inner thin disc. Our new results, combined with data on abundances in 
the very central part of our Galaxy taken from literature, show that 
iron, magnesium, silicon, sulfur, calcium and titanium LTE abundance 
radial distributions, as well as NLTE distribution of oxygen reveal 
a plateau-like structure or even positive abundance gradient in the 
region extending from the Galactic center to about 5 kpc. 
\end{abstract}


\section{Introduction}
In our series of papers (\citealt{AND02}, \citealt{An02b}, 
\citealt{An02c}, \citealt{AND04}, \citealt{Lu03}, \citealt{Lu06}, 
\citealt{luc11}) we presented the radial elemental distributions 
over a wide interval of Galactocentric distances based on high-resolution 
spectroscopic data of a large sample of Cepheids (thin disc population). 
Indeed, we succeeded in covering more than 10 kpc in radial direction from 
about 4 to 16 kpc. A gradual increase of elemental abundance as distance 
decreases was confirmed for many chemical elements. In particular, 
\citet{An02b} used a limited number of Cepheid stars to find the properties 
of the metallicity distribution in the inner part of the Galactic disc. 
The number of studied stars with Galactocentric distances less than 
6 kpc at that time was only five, but almost all chemical elements 
investigated in those stars showed clearly increased abundances 
comparing to the Cepheids from the solar vicinity (R$_{\odot} \approx 8.0$ 
kpc).  

Recently \citet{Gen13} presented probably the most comprehensive 
compilation of data on metallicity distribution obtained from Cepheids 
in the Galactic inner disc (with Galactocentric distances $\geq$ 4.6 kpc). 
That paper contains the corresponding references on relevant studies, 
and will not be repeated here.

The increased sophistication of observing techniques in the
recent decade allows us to obtain high-resolution stellar spectra near the 
very center of our Galaxy. This is often achieved by using the far- 
and near-IR domains of the electromagnetic spectrum where absorption is much less than in 
the visual region. \citet{Ryd15} summarize all the results for 
Galactic center abundances obtained up to now. 

The general picture of the metallicity distribution in the inner part 
of the Galactic disc can be drawn from all of these studies: the metallicity 
of the thin disc gradually increases toward the Galactic center, and reaches 
about [Fe/H] $ \approx +0.5$ at R$_{\rm G}  \approx 4$ kpc. The region of 
the thin disc between this distance and the immediate Galactic center still 
remains poorly sampled, and the spatial properties of the metallicity 
distribution in this domain are not known. Filling in this gap would have 
a certain interest for the Galactic chemical evolution models. 

In this paper we present abundances for stars which are mainly 
situated in the region 5-7 kpc, with one Cepheid located in the 
critical region at about 3 kpc. Observational details are given 
in the next Section. Stellar atmosphere parameters relevant to 
our study are brifely discussed in Section 3. Abundance measurements 
and their radial distribution are presented in Section 4 and 5, 
respectively. Section 6 summarizes our results and conclusions.

\section{Observations and data reduction}

The observations were carried out with 3.6 m Canada-France-Hawaii Telescope 
(CFHT), the Very Large Telescope (VLT) Unit 2 Kueyen, and the 2.2 m 
MPG Telescope.

\emph{CFHT}. The observations were carried out 
under the queue observing mode using the fiber-fed ESPaDOnS echelle 
spectrograph equipped with an e2v 2048 x 4608 CCD (binned 1 x 1). The 
resolving power provided by this combination was about 80000 and the 
spectral range extended from 370 to 1050 nm. The spectra were processed 
by the CFHT ESPaDOnS pipeline. The estimated S/N ratio at the continuum level 
depends upon the wavelength interval, and varies for each star in the range 
from about  80 to 100.

\emph{VLT}. For data obtained with the VLT (observing run 089.D-0489),
the red arm of the UVES cross-dispersed echelle spectrograph was used.
With the red arm, the wavelength region between 420 nm and 1100 nm can
be scanned. The grating angle was set at 22.668 degrees thus allowing
to cover a wavelength interval between 479 nm and 681 nm. The light
is dispersed over two CCD chips, with a gap between 576 nm and 583 nm.
The slit was set at 1$^{\prime\prime}$ width and 12$^{\prime\prime}$
length, yielding a resolution of 38700. The median SNR for the upper
and lower chip spectra are $\approx$50 and $\approx$14, respectively but
with a relatively large range: from $\approx$32 to $\approx$56 for 
the upper, and from $\approx$11 to $\approx$40 for the lower chip.

\emph{ESO/MPGT.} For data taken with the ESO/MPG 2.2 m telescope, the 
FEROS spectrograph was used \citep{Ka99}. The instrument has a 
wavelength coverage from $\approx$350 nm to $\approx$920 nm, and 
a resolving power of 48000. The typical SNR ratio of our spectra 
is $\approx90$. Each night a broad-line B star was observed with 
a SNR exceeding that of the program stars, in order to enable 
cancellation of the telluric lines when necessary.

Table \ref{Par} contains details on our observations of Cepheids
for this study, as well as some information about the physical properties of 
the Cepheids themselves (see next Section).

The processing of spectra (continuum level determination, equivalent widths 
measurements etc.) was carried out by using the DECH20 software package 
\citep{Gal92}. The line-list is the same as used in our previous 
studies (see \citealt{Kov99}).

\begin{table*}
\small
\caption{Observations and parameters of the investigated stars}
\label{Par}
\begin{tabular}{rrcrcccccc}
\hline
 Star           &  P, d      & JD(start) 245..& V     &E(B-V)&Telescope& T$_{\rm eff}$, K & 
$\log~g$&V$_{\rm t}$,km~s$^{-1}$&R$_{\rm G}$, kpc \\
\hline                                           
V340 Ara        & 20.8090000 & 4186.72337651  & 10.164&0.546 &  MPGT   & 5318 & 1.15 & 4.2 &  4.34 \\
SS CMa          & 12.3610001 & 6289.79272864  &  9.915&0.549 &  MPGT   & 5380 & 1.50 & 4.5 &  9.71 \\
TV CMA          &  4.6700101 & 6289.78263332  & 10.582&0.583 &  MPGT   & 5754 & 2.10 & 3.7 &  9.56 \\
   AG Cru       &  3.83728   & 5966.84795300  &  8.225&0.212 &  MPGT   & 5730 & 2.15 & 4.6 &  7.35 \\
 X Cru          &  6.2199702 & 5966.87123636  &  8.404&0.272 &  MPGT   & 5394 & 1.70 & 4.5 &  7.20 \\
V1033 Cyg       &  4.9375119 & 6524.95738426  & 13.027&1.067 &  CFHT   & 5819 & 2.10 & 3.4 &  7.46 \\
EK Pup          &  2.6259401 & 6289.79920932  & 10.664&0.328 &  MPGT   & 6434 & 2.30 & 4.0 &  9.73 \\
WY Pup          &  5.2508001 & 6289.82627777  & 10.569&0.270 &  MPGT   & 6537 & 2.30 & 5.0 & 10.26 \\
KQ Sco          & 28.6896000 & 6141.67867798  &  9.807&0.869 &  MPGT   & 5697 & 2.00 & 6.2 &  5.41 \\
V470 Sco        & 16.2614994 & 6518.72862269  & 11.069&1.517 &  CFHT   & 5906 & 1.70 & 4.9 &  6.62 \\
    X Sct       &  4.1980700 & 6524.80085648  & 10.006&0.619 &  CFHT   & 5811 & 2.15 & 4.7 &  6.46 \\
    Y Sct       & 10.3414831 & 6524.81630787  &  9.628&0.823 &  CFHT   & 5145 & 1.60 & 3.8 &  6.55 \\
                &            & 6092.89682642  &       &      &  VLT    & 5643 & 2.00 & 6.0 &       \\
   BX Sct       &  6.4113302 & 6524.86835648  & 12.241&1.318 &  CFHT   & 5599 & 2.00 & 3.6 &  6.38 \\
   CK Sct       &  7.4152198 & 6518.82194444  & 10.590&0.795 &  CFHT   & 5735 & 2.00 & 4.3 &  6.13 \\
   CM Sct       &  3.9169769 & 6518.87042824  & 11.106&0.771 &  CFHT   & 5679 & 2.00 & 3.6 &  6.23 \\
   CN Sct       &  9.9923000 & 6524.83060185  & 12.478&1.267 &  CFHT   & 5456 & 1.80 & 5.0 &  5.64 \\
   RU Sct       & 19.7006207 & 6524.82368056  &  9.466&0.957 &  CFHT   & 5180 & 1.50 & 4.7 &  6.52 \\
   SU Sct       &  1.4679780 & 6524.88760417  & 13.663&0.459 &  CFHT   & 5975 & 2.80 & 3.3 &  3.22 \\
                &            & 6069.88148484  &       &      &  VLT    & 6190 & 2.60 & 4.5 &       \\
   TY Sct       & 11.05302   & 6518.85549769  & 10.831&1.014 &  CFHT   & 5487 & 1.90 & 5.5 &  6.16 \\
                &            & 6092.90857267  &       &      &  VLT    & 5811 & 1.80 & 4.8 &       \\
   AA Ser       & 17.1411991 & 6518.83685185  & 12.228&1.603 &  CFHT   & 5388 & 1.50 & 4.2 &  6.30 \\
   CR Ser       &  5.3014102 & 6524.78538194  & 10.842&1.011 &  CFHT   & 6169 & 2.15 & 4.2 &  6.57 \\
   AY Sgr       &  6.5695901 & 6518.80657407  & 10.549&0.919 &  CFHT   & 5714 & 1.90 & 3.7 &  6.36 \\
 V773 Sgr       &  5.7484450 & 6524.74738426  & 12.389&1.620 &  CFHT   & 5467 & 1.50 & 3.9 &  6.73 \\
V1954 Sgr       &  6.1794491 & 6518.79149306  & 10.831&0.875 &  CFHT   & 5501 & 1.80 & 3.8 &  6.07 \\
V5567 Sgr       &  9.7616    & 6147.73398282  & 10.550&0.990 &  VLT    & 5868 & 2.00 & 4.7 &  6.26 \\
ASAS 171305-4323&  9.555323  & 6124.72436291  & 12.645&1.630 &  VLT    & 6153 & 2.40 & 5.4 &  5.96 \\
   BD -04 4569  & 13.83873   & 6124.81107798  & 10.026&1.300 &  VLT    & 5620 & 2.30 & 7.0 &  7.00 \\
\hline
\end{tabular}

Remark: Reddening values are from \citet{Fo07} or 
\citet{Fer95} with the systematic correction of 
\citet{Fo07} applied (factor 0.952). These reddening values 
are on the E(B-V) scale by \citet{Lan07}. Note that 
for some stars the reddening values were found using a special 
method (see Sect. 4).
\end{table*}

\section{Stellar atmosphere parameters}

The effective temperature (T$_{\rm eff}$) for the stars in our program was 
estimated by calibrating the ratios of the central depths of the lines with 
different potentials of the lower levels (see \citealt{Kov07}). The surface 
gravity values were computed using the iron ionization balance. The 
microturbulent velocity was found by avoiding any dependence between 
the iron abundance as produced by individual Fe~II lines and their 
equivalent widths. The resulting atmosphere parameters for the 
studied stars are listed in Table \ref{Par}.

\section{Spectroscopic analysis}

All abundances listed in Tables \ref{Ab1} and \ref{Ab2} (with an exception 
for oxygen, sulfur and barium) were derived in the LTE approximation in order to 
be consistent with our previous abundance determinations in Cepheids. 
For this we used the ATLAS9 code \citep{Kur92} to generate the 
appropriate atmosphere models, and Kurucz's WIDTH9 code to analyze 
the equivalent widths. The reference solar abundances were taken
from \citet{Gr98}.

\begin{table*}
\caption{The abundances of the investigated stars (C--Mn)}
\label{Ab1}
\small
\begin{tabular}{r@{\extracolsep{-0.2mm}}rrrrrrrrrrrrrrr}
\hline
     Star              &    C  &   N  &   O   &   Na &   Mg &   Al &   Si &   S  &$\rm S_{LTE}$&   Ca &   Sc &  Ti  &  V   &  Cr  &   Mn  \\
\hline                                                                               
    V340 Ara           &   0.23&  0.89&  0.36 &  0.55&  0.21&  0.40&  0.31&  0.30 &  0.48&  0.16&  0.30&  0.23&  0.21&  0.30&  0.30 \\
      SS CMa           & --0.19&  0.68&--0.01 &  0.17&--0.18&  0.05&--0.03&  0.03 &--0.13&--0.20& -0.16&--0.06&--0.12&--0.14&--0.16 \\
      TV CMa           & --0.17&  0.46&--0.08 &  0.14&--0.03&--0.01&  0.06&--0.02 &  0.12&--0.12&  0.02&  0.00&--0.03&--0.04&--0.15 \\
      AG Cru           & --0.14&  0.53&--0.04 &  0.20&--0.09&  0.08&  0.02&--0.01 &  0.12&--0.07& -0.03&--0.02&--0.03&--0.01&--0.11 \\
       X Cru           & --0.05&  0.18&  0.06 &  0.24&--0.19&  0.10&  0.05&  0.09 &  0.15&--0.03& -0.05&--0.04&--0.06&--0.04&--0.02 \\
   V1033 Cyg           &   0.05&--0.09&  0.01 &--0.17&--0.05&  0.04&  0.06&  0.02 &  0.16&  0.03&  0.02&  0.05&--0.02&--0.01&--0.13 \\
      EK Pup           & --0.41&  0.31&--0.15 &--0.09&--0.26&--0.20&--0.07&--0.18 &--0.16&--0.29& -0.19&--0.21&--0.14&--0.27&--0.28 \\
      WY Pup           & --0.42&--0.09&--0.17 &--0.20&--0.27&  ... &--0.16&--0.28 &--0.29&--0.38& -0.41&--0.37&--0.21&--0.41&--0.37 \\
      KQ Sco           &   0.10&  0.71&  0.36 &  0.44&  0.34&  0.26&  0.26&  0.30 &  0.40&  0.17&  0.35&  0.27&  0.29&  0.36&  0.27 \\
    V470 Sco           &   0.12&  0.66&  0.18 &  0.57&  0.26&  0.34&  0.30&  0.15 &  0.45&  0.26&  0.33&  0.32&  0.30&  0.31&  0.22 \\
       X Sct           & --0.08&  0.35&--0.02 &  0.24&  0.00&  0.05&  0.04&--0.01 &  0.11&  0.20&  0.04&  0.10&  0.05&  0.10&--0.11 \\
       Y Sct           & --0.12&  0.75&  0.25 &  0.28&--0.03&  0.15&  0.10&  0.16 &  0.16&  0.10&  0.11&  0.08&  0.05&  0.08&  0.04 \\
      BX Sct           & --0.07&  0.43&  0.12 &  0.24&  0.01&  0.14&  0.10&  0.14 &  0.23&  0.11&  0.09&  0.10&  0.03&  0.05&--0.01 \\
      CK Sct           & --0.02&  0.41&  0.15 &  0.15&--0.06&  0.07&  0.10&  0.14 &  0.22&  0.10&  0.12&  0.13&  0.04&  0.07&--0.06 \\
      CM Sct           & --0.16&  0.28&  0.13 &  0.22&--0.02&  0.04&  0.07&--0.01 &  0.13&  0.07&  0.07&  0.03&  0.02&  0.04&--0.08 \\
      CN Sct           &   0.20&  0.77&  0.29 &  0.38&  0.35&  0.20&  0.28&  0.25 &  0.44&  0.14&  ... &  0.23&  0.20&  0.13&  0.15 \\
      RU Sct           &   0.06&  0.66&  0.15 &  0.20&--0.09&  0.09&  0.10&  0.16 &  0.30&  0.05&  0.08&  0.02&  0.01&  0.04&  0.04 \\
      SU Sct           &   0.31&  0.62&  0.35 &  0.10&  0.19&--0.04&  0.17&  0.29 &  0.36&--0.01&  0.11&  0.09&  0.40&  0.26&  0.22 \\
      TY Sct           &   0.18&  0.60&  0.29 &  0.40&  0.16&  0.26&  0.21&  0.27 &  0.39&  0.18&  0.30&  0.23&  0.23&  0.15&  0.15\\
      AA Ser           &   0.31&  0.90&  0.40 &  0.48&  0.27&  0.34&  0.30&  0.34 &  0.58&  0.32&  0.35&  0.18&  0.28&  0.33&  0.25 \\
      CR Ser           & --0.18&  0.53&  0.07 &  0.26&  0.12&  0.08&  0.14&  0.04 &  0.24&  0.02&  0.07&  0.12&  0.10&  0.13&--0.01 \\
      AY Sgr           & --0.01&  0.45&  0.18 &  0.35&  0.18&  0.18&  0.18&  0.16 &  0.32&  0.15&  0.17&  0.12&  0.12&  0.16&  0.10 \\
    V773 Sgr           & --0.13&  0.32&  0.02 &  0.29&  0.11&  0.19&  0.18&  0.01 &  0.18&  0.17&  0.02&  0.13&  0.07&  0.10&  0.12 \\
   V1954 Sgr           &   0.13&  0.65&  0.16 &  0.56&  0.19&  0.32&  0.30&  0.20 &  0.43&  0.25&  0.23&  0.21&  0.21&  0.26&  0.28 \\
   V5567 Sgr           & --0.17&  ... &  0.18 &  0.19&--0.16&  0.09&  0.05&  ...  &  0.08&  0.03&  0.06&  0.07&  0.01&  0.05&--0.11 \\
   ASAS 171305-4323    &   0.10&  ... &  0.24 &  0.51&  0.11&  0.26&  0.28&  ...  &  0.41&  0.31&  0.46&  0.29&  0.32&  0.34&  0.20 \\
    BD-04 4569         &   0.04&  ... &  ...  &  0.05&  ... &  0.15&  0.03&  ...  &  0.14&  0.02&  ... &  0.10&  0.25&  0.28&--0.01 \\
\hline
\end{tabular}
 \end{table*}

\begin{table*}
\caption{The abundances of the investigated stars (Fe--Gd)}
\label{Ab2}
\small
\begin{tabular}{r@{\extracolsep{-0.1mm}}rrrrrrrrrrrrrrr}
\hline
       Star            &    Fe  &   Co &   Ni &   Cu &  Zn  &  Y   &  Zr  & Ba  & La  &  Ce  &  Pr  &  Nd  &   Sm &  Eu  &   Gd    \\
\hline                                                                           
    V340 Ara           &    0.32&  0.18&  0.35&  0.28&  0.36&  0.32&  0.21&  0.17 &  0.18&   0.00&  0.09&  0.03&  0.43&  0.28&  0.23   \\
      SS CMa           &  --0.02&--0.28&--0.12&--0.23&--0.17&  0.11&  0.18&  0.27 &  0.18& --0.01&--0.04&--0.02&  ... &  0.01&  ...    \\
      TV CMa           &    0.03&--0.18&--0.04&--0.12&--0.29&  0.20&  0.22&  0.20 &  0.14&   0.15&--0.06&  0.01&  ... &  0.03&  ...    \\
      AG Cru           &    0.01&--0.22&--0.02&--0.08&  0.04&  0.17&--0.02&--0.15 &  0.17& --0.06&--0.30&--0.10&   ...&  0.10&  ...    \\
       X Cru           &    0.05&--0.21&--0.03&--0.13&  0.08&  0.17&  0.13&  0.00 &  0.19&   0.06&--0.24&--0.05&  ... &  0.10&  ...    \\
   V1033 Cyg           &    0.04&  0.02&  0.01&--0.04&  0.11&  0.16&  0.09&  0.14 &--0.03&   0.12&  0.05&  0.09&  ... &  0.16&  ...    \\
      EK Pup           &  --0.18&--0.22&--0.21&--0.40&--0.47&--0.05&--0.02&--0.05 &  ... & --0.13&--0.07&--0.10&  ... &  0.06&  ...    \\
      WY Pup           &  --0.35&  ... &--0.39&--0.26&--0.43&--0.22&--0.05&  0.02 &  ... & --0.04&  ... &  0.02&  ... &  ... &  ...    \\
      KQ Sco           &    0.35&  0.21&  0.35&  0.30&  0.34&  0.37&  0.39&  0.20 &  0.26&   0.05&  0.20&  0.14&  0.48&  0.38&  0.44   \\
    V470 Sco           &    0.30&  0.26&  0.32&  0.13&  0.19&  0.36&  0.37&  0.09 &  0.29&   0.09&  0.08&  0.18&  0.17&  0.28&  0.32   \\
       X Sct           &    0.06&--0.06&  0.04&  0.03&  0.14&  0.15&  0.13&  0.08 &  0.19&   0.09&  0.06&  0.11&  0.32&  0.30&  0.32   \\
       Y Sct           &    0.13&  0.00&  0.10&--0.06&  0.29&  0.26&  0.10&  0.25 &  0.20&   0.08&  0.02&  0.06&  0.35&  0.20&  0.17   \\
      BX Sct           &    0.13&  0.03&  0.11&  0.03&  0.16&  0.34&  ... &  0.23 &  0.31&   0.14&--0.05&  0.18&  ... &  0.20&  ...    \\
      CK Sct           &    0.12&  0.01&  0.09&--0.04&  0.10&  0.29&  0.07&  0.19 &  0.17&   0.09&--0.04&  0.06&  ... &  0.14&  ...    \\
      CM Sct           &    0.07&--0.05&  0.04&--0.11&  0.02&  0.22&  0.12&  0.17 &--0.09&   0.06&--0.07&  0.05&  ... &  0.10&  ...    \\
      CN Sct           &    0.28&  0.13&  0.25&  0.20&  ... &  0.37&  0.13&  0.20 &  0.28&   0.18&  ... &  0.43&  ... &  0.29&  ...    \\
      RU Sct           &    0.15&--0.04&  0.06&--0.20&  0.25&  0.27&  0.19&  0.14 &  0.06&   0.06&  0.03&  0.15&  0.25&  0.20&  0.17   \\
      SU Sct           &    0.27&  0.44&  0.20&  0.29&  0.05&  0.15&  0.32&  0.25 &  0.49&   0.48&  ... &  0.09&  ... &  0.34&  ...    \\
      TY Sct           &    0.25&  0.18&  0.30&  0.17&  0.39&  0.30&  0.34&  0.18 &  0.18&   0.08&  ... &  0.08&   ...&  0.32&  ...    \\
      AA Ser           &    0.33&  0.18&  0.35&  0.29&  ... &  0.36&  0.39&  0.17 &  0.25&   0.06&  0.07&  0.04&  0.15&  0.33&  0.33   \\
      CR Ser           &    0.11&  0.05&  0.10&  0.04&  0.09&  0.27&  0.20&  0.14 &  0.07&   0.21&  0.09&  0.18&  0.10&  0.22&  0.24   \\
      AY Sgr           &    0.20&  0.05&  0.16&  0.05&  0.25&  0.31&  0.27&  0.32 &  0.13&   0.20&  0.09&  0.14&  0.14&  0.17&  0.32   \\
    V773 Sgr           &    0.11&--0.01&  0.13&  0.04&  0.16&  0.15&  0.14&  0.12 &  0.17&   0.02&  0.03&  0.16&  0.38&  0.21&  ...    \\
   V1954 Sgr           &    0.29&  0.17&  0.28&  0.17&  ... &  0.28&  0.29&  0.17 &  0.15&   0.14&  0.03&  0.09&  0.33&  0.32&  ...    \\
   V5567 Sgr           &    0.07&--0.05&  0.00&--0.14&--0.21&  0.25&  0.15&  0.05 &  0.02&   0.18&--0.13&  0.06&  0.03&  0.18&  0.05   \\
ASAS 171305-4323       &    0.32&  0.27&  0.29&  ... &  0.33&  0.42&  0.32&--0.05 &  0.32&   0.24&  0.12&  0.15&  0.31&  0.35&  0.26   \\
  BD-04 4569           &    0.08&  ... &  0.08&  ... &  ... &  0.18&  0.36&  0.05 &  0.45&   0.22&  ... &  0.18&  ... &  0.41&  ...    \\
\hline
\end{tabular}
\end{table*}

Expected errors in abundances caused by variations in the 
fundamental stellar parameters
are presented in Table \ref{Aberr} for our critical object
SU~Sct. The total abundance error was obtained by increasing the
temperature by 100\,K, surface gravity by 0.2 dex and microturbulent 
velocity by 0.5 km~s$^{-1}$. Decreasing these parameters by similar
values will produce almost the same errors.

Our sample includes several stars in common with the 
studies of  \citet{Gen14} and \citet{ll11}. 
In Table \ref{comptab} we compare our iron 
abundance measurements with those of the above 
studies. Generally, there is reasonable agreement between 
these independent results.

An exception in this paper was made for oxygen and barium whose NLTE 
abundances for a large sample of Cepheids were previously derived 
by us (\citealt{luc13}, \citealt{Kor14}, \citealt{AND13}, 
\citealt{AND14}). In those papers the reader can find all the 
details concerning the method of the NLTE calculations and applied 
atomic models of oxygen and barium. For sulfur we present both LTE 
and NLTE abundances. The NLTE sulfur atomic model used for the NLTE 
analysis was described by \citet{Kor09}.

In the next section we present the radial abundance distributions for 
some chemical elements. The distances for the stars of the present sample 
were found in the same way as in \citet{AND02}. For three 
stars of our program the E(B-V) values, which are necessary to estimate 
the heliocentric distances, are not available in the literature 
(the stars V5567 Sgr,  ASAS 171305-4323, and BD-04 4569).
Therefore they were derived by us using the correlation between the 
diffuse interstellar band (DIB) equivalent widths measured in their 
spectra and E(B-V) indices as proposed by \citet{Fr11}. For this 
we used the interstellar features at 5780.5, 5797.1 and 6613.6 \AA. 
To remove the stellar lines from the DIB region before the equivalent 
width measurement, we used a synthetic spectrum technique. The accuracy 
of the E(B-V) values determined by this method is approximately 10-15\%.

For the star SU Sct, which has the smallest Galactocentric distance 
in our sample, we compared our estimate of the E(B-V) to the literature 
value provided by \citet{Ng12}, who applied the Wesenheit function  
in deriving distances to Galactic Cepheids. According to this method, 
the heliocentric distance of SU Sct is 6.62 kpc, that gives R$_{\rm G} 
= 3.04$ kpc, in good agreement with our value listed in Table \ref{Par}.

\section{Results and discussion}

\begin{figure}
\resizebox{\hsize}{!}           	
{\includegraphics {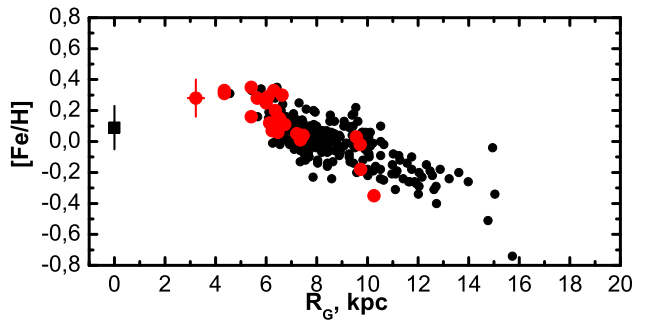}}
\caption[]{LTE iron abundance distribution in Galactic disc. Black 
circles - our previous determinations published in a series of papers 
mentioned in the Introduction. Red circles - present sample. For SU Sct 
we show the typical for Cepheids the iron abundance uncertainty 
($2\sigma$), and 10\% uncertainty in the distance. The Galactic center 
iron abundance is a mean value calculated from the individual star 
abundances published by \citet{Car00}, \citet{Ram00}, \citet{Cun07},
\citet{Dav09} and \citet{Ryd15}. 
The $2\sigma$ interval is showed for this data point.}
\label {FeRg}
\end{figure}

In Fig. \ref{FeRg} we show the iron abundance distribution of Galactic 
Cepheids from our previous studies and the present sample together with 
the Galactic center abundances. The Cepheids of our program
cover a similar pulsation period range as the stars included in our previous 
samples; thus they share a similar age range. Following the period-age relation 
for classical Cepheids as proposed by \citet{Bono05}, the age range of the Cepheids 
in our sample is roughly 20-70 Myr. The average iron abundance  at the 
Galactic Center from literature presented in Fig. \ref{FeRg} was mostly determined from 
stars covering a larger age range, that is from members of the Arches/Quintuplet clusters ($\sim$ 3 -9 Myr) 
to field stars of $\sim$ 1Gyr. For iron and other elements, 
the absolute abundances derived by other authors were normalized to the solar 
abundances from \citet{Gr98}. 

As explained earlier, one of our stars (SU Sct) in the critical region 
between the central part and 5 kpc gives some reason to believe that the 
iron content in the Galactic thin disc young population stars (in Cepheids, 
in particular) does not exceed the value of about +0.4 dex if we use our 
homogeneous data, or +0.5 when using the compilation data of \citet{Gen13}. 
After reaching the value of  [Fe/H] = +0.4 dex at about 5 kpc 
the metallicity tends to slightly decrease in the direction of the 
Galactic center.

\begin{figure}
\includegraphics[width=8.5cm]{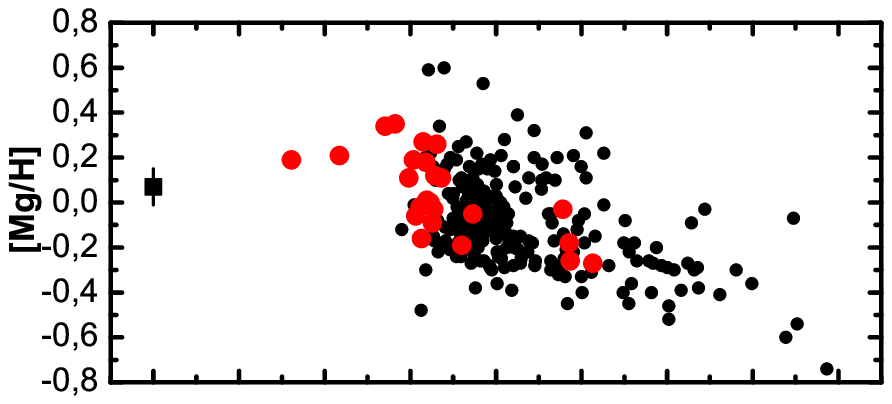}
\includegraphics[width=8.5cm]{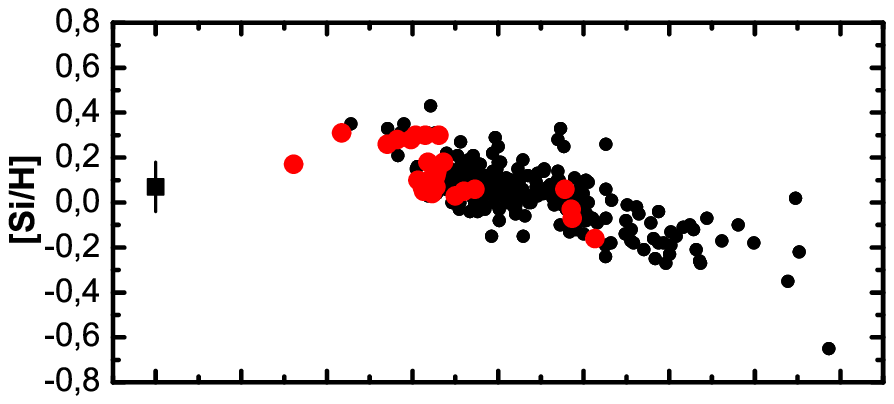}
\includegraphics[width=8.5cm]{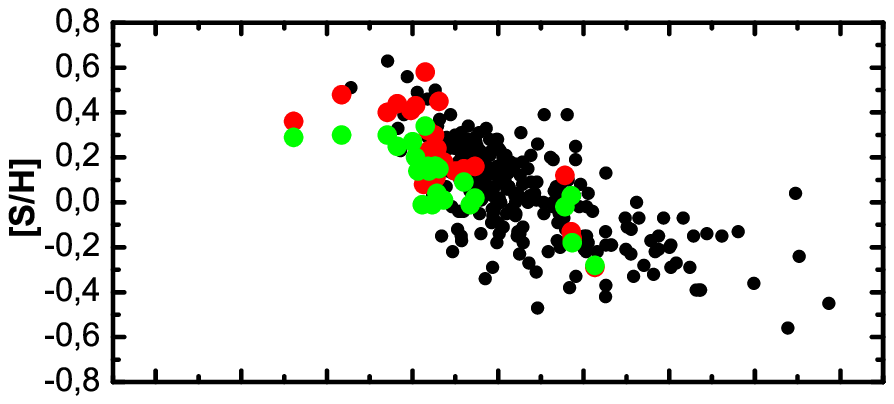}
\includegraphics[width=8.5cm]{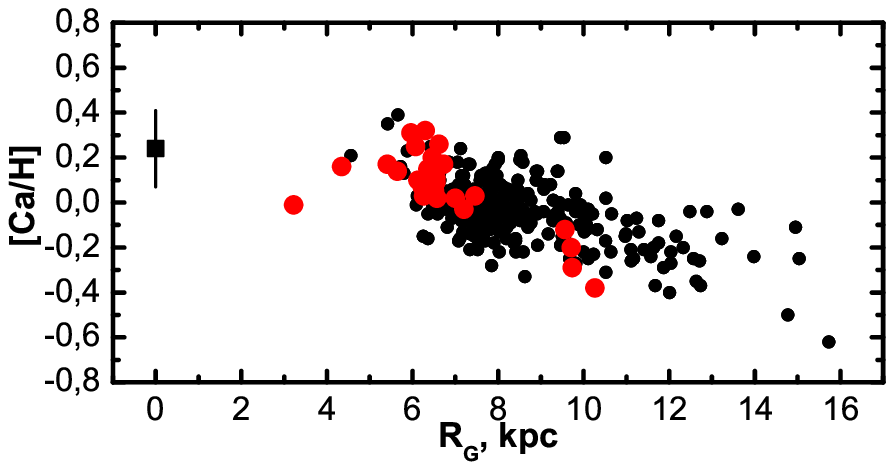}
\includegraphics[width=8.5cm]{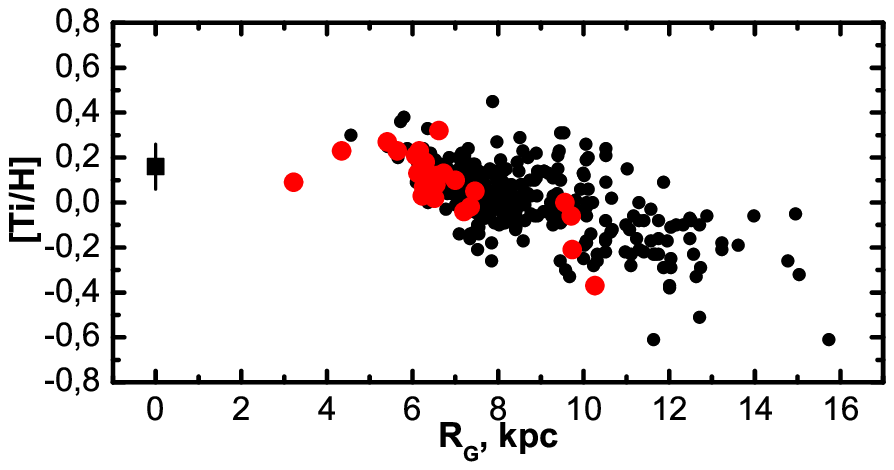}
\caption[]{The same as Fig.1 but for magnesium, silicon, sulfur, 
calcium and titanium. Green circles in the sulfur plot - our NLTE 
abundances for the stars of present sample. The Galactic center 
abundance data are from \citet{Cun07}, \citet{Dav09} 
and \citet{Ryd15}. The Galactic center abundance of sulfur 
is not available, but we show this element together with other 
$\alpha$-element abundance distributions.}
\label {alRg}
\end{figure}

Similar trends can also be seen for such elements as magnesium, silicon, 
sulfur, calcium and titanium, whose abundances in the Galactic center 
nowadays are known (Fig. \ref{alRg}). Despite the fact that abundances 
of these elements were derived in LTE, we believe that the general 
tendency noticed for iron, and also seen for these $\alpha$-elements, 
will not change qualitatively after the applying the NLTE corrections 
(in particular, this conclusion is supported by our result for 
sulfur, where we show both LTE and NLTE abundances for the present 
sample of stars).

\begin{figure}
\includegraphics[width=8.5cm]{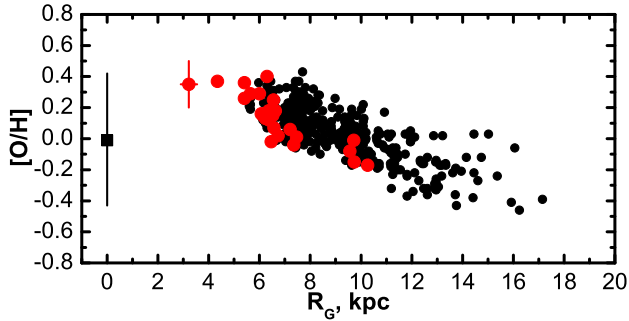}
\caption[]{The NLTE abundance distribution of oxygen vs. Galactocentric 
distance. Black circles - \citet{luc13} and \citet{Kor14}, 
red circles - present sample of Cepheids. The Galactic center abundance 
is the mean value based on the data from \citet{Car00}, \citet{Cun07} 
and \citet{Dav09}.}
\label {ORg}
\end{figure}

As one can see, the iron and $\alpha$-elements demonstrate a clear 
plateau in their abundance distributions, or even positive gradient 
value in the region from the Galactic center to about 5 kpc.  This 
is also clearly seen in our fully NLTE abundance distribution of the 
oxygen (see Fig. \ref{ORg}). Similar to iron, the highest oxygen abundance 
in the thin disc is achieved at about 4-5 kpc at a value of about 
+0.4 dex. Figures \ref{alRg} and \ref{ORg} show that there is no 
overabundance of the oxygen and $\alpha$-elements with respect to 
iron, in agreement with what is expected for thin disc stars.

\begin{figure}
\includegraphics[width=8.5cm]{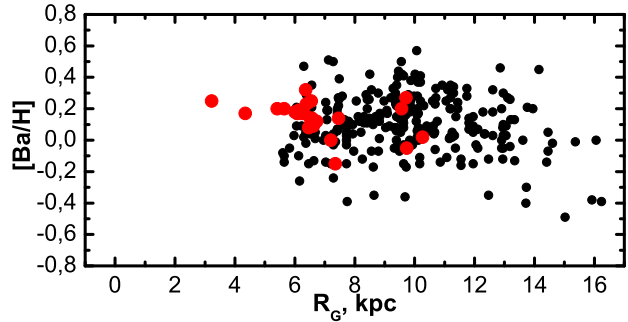}
\caption[]{The NLTE abundance distribution of barium vs. Galactocentric 
distance. Black circles - \citet{AND13}, \citet{AND14}, 
red circles - present sample of Cepheids. The Galactic center 
abundance of this element is not available.}
\label {BaRg}
\end{figure}

It is also interesting to note that in contrast with elements 
discussed above, barium, which is mainly produced by low mass AGB 
stars, does not display any remarkable trend with the Galactocentric 
distance neither in the outer or in the inner part of the Galactic 
disc, as shown in Fig. \ref{BaRg}.

\begin{figure}
\includegraphics[width=8.5cm]{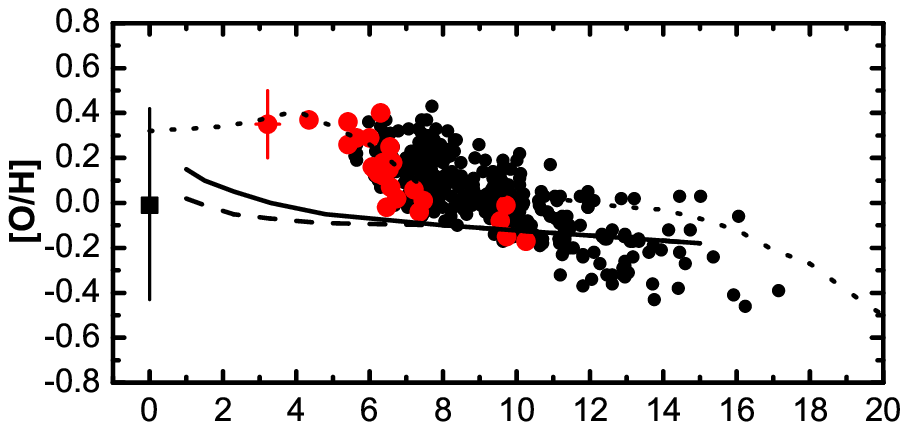}
\includegraphics[width=8.5cm]{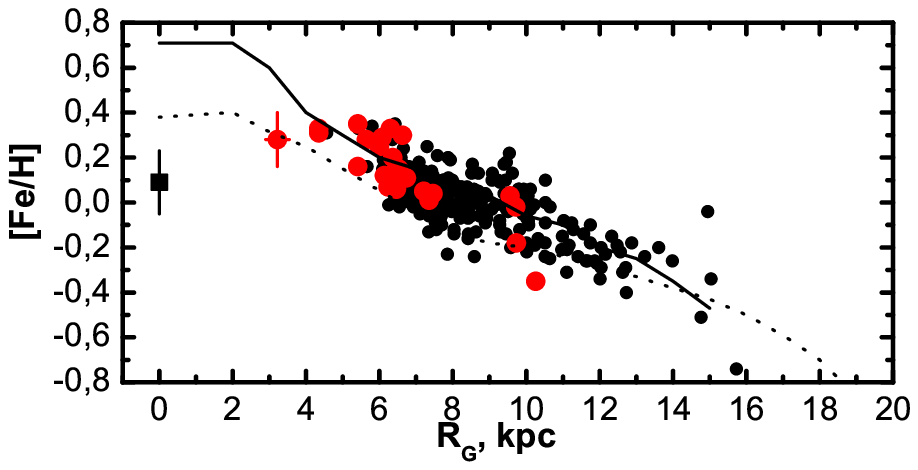}
\caption[]{Our abundance distributions of oxygen and iron superimposed 
with theoretical curves for oxygen from \citet{Cav14} - dotted 
line, and \citet{Fu13} - the model with radial gas inflow - solid 
line, and without it - dashed line, and for iron from \citet{Cav14}
- dotted line, and \citet{Min13} - solid line. Note that we re-scaled
the absolute oxygen abundances of \citet{Cav14} assuming that their ``solar" oxygen abundance (i.e. abundance at R$_{\rm G} \approx$ 8 kpc)
12+log(O/H) = 8.4.}
\label {OFe}
\end{figure}

\citet{Br88} derived the mean radial distribution of molecular clouds in 
Galactic disc, and found that the highest density of interstellar gas 
is observed in the Galactocentric distance range of 4 to 8 kpc with a 
maximum value at about 5-6 kpc. At the distance of about 3-4 kpc there 
is an abrupt drop in the gas surface density. The same result follows 
from the papers of \citet{San84} and \citet{Dam93}. If a lower gas density 
causes a decrease in the star formation rate in the region with Galactocentric 
distance less than 4 kpc, then the chemical enrichment of interstellar matter 
from SNe will not be effective in this region. This should be 
imprinted on the derived abundances of the young population stars 
within the Galactic bar, whose influence on the chemical properties 
of the inner part of our Galaxy must be strong. Existing theoretical models 
that deal with the inner part of the Galaxy do not reproduce 
these observed trends in the central part (0-6 kpc) quite correctly, 
although some of them take into account the radial gas flow, as 
well as the bar influence on the chemistry of the inner disc. 
Nevertheless, the general improvement seen with those models
seems to be clear. This is demonstrated by the oxygen and iron 
plots in Fig. \ref{OFe}.

\begin{table}
\caption{Abundance uncertainties due to stellar atmospheric parameters.
 SU~Sct (VLT,  T$_{\rm eff}$=6190 \,K, $\log~g$=2.6, V$_{\rm t}$=4.5 km~s$^{-1}$, 
[Fe/H] = +0.26).}
\label{Aberr}
\begin{tabular}{rrrrr}
\hline
Species & $\Delta$ T$_{\rm eff}$& 
$\Delta$ $\log~g$ & $\Delta$ V$_{\rm t}$ & Total error\\
& +100\,K  & +0.2 & +0.5 km~s$^{-1}$ & \\
\hline
 6.00 & --0.05 &   0.06 & --0.02&  0.08  \\
11.00 &   0.05 & --0.01 & --0.03&  0.06  \\
12.00 &   0.05 & --0.01 & --0.03&  0.06  \\
13.00 &   0.05 & --0.01 & --0.01&  0.05  \\
14.00 &   0.04 & --0.01 & --0.02&  0.05  \\
14.01 & --0.08 &   0.06 & --0.02&  0.10  \\
16.00 & --0.03 &   0.04 & --0.03&  0.06  \\
20.00 &   0.06 & --0.01 & --0.04&  0.07  \\
21.01 &   0.02 &   0.08 & --0.04&  0.09  \\
22.00 &   0.08 & --0.01 & --0.01&  0.08  \\
22.01 &   0.02 &   0.08 & --0.05&  0.10  \\
23.00 &   0.09 & --0.02 & --0.01&  0.09  \\
23.01 &   0.02 &   0.08 & --0.01&  0.08  \\
24.00 &   0.07 & --0.01 & --0.01&  0.07  \\
24.01 & --0.01 &   0.08 & --0.05&  0.09  \\
25.00 &   0.07 & --0.01 & --0.02&  0.07  \\
26.00 &   0.07 & --0.01 & --0.02&  0.07  \\
26.01 & --0.01 &   0.07 & --0.05&  0.09  \\
27.00 &   0.07 & --0.02 & --0.01&  0.07  \\
28.00 &   0.07 & --0.01 & --0.02&  0.07  \\
29.00 &   0.08 & --0.02 & --0.03&  0.09  \\
30.00 &   0.07 &   0.00 & --0.05&  0.09  \\
39.01 &   0.03 &   0.07 & --0.02&  0.08  \\
40.01 &   0.03 &   0.07 & --0.01&  0.08  \\
57.01 &   0.04 &   0.07 & --0.01&  0.08  \\
58.01 &   0.04 &   0.07 & --0.01&  0.08  \\
60.01 &   0.05 &   0.07 &   0.00&  0.09  \\
63.01 &   0.03 &   0.07 & --0.02&  0.08  \\
 \hline
\end{tabular}
\end{table}

\begin{table}
\caption{Comparison of the iron abundance results.}
\label{comptab}
\begin{center}
\begin{tabular}{rrcc}
\hline
 Star         &This paper& GET & LL\\
\hline
     V340 Ara &    0.32&  0.33$\pm$ 0.09&      \\
       SS CMa &  --0.02&                & 0.07 \\
       TV CMa &    0.03&                & 0.14 \\
        X Cru &    0.05&                & 0.15 \\
       AG Cru &    0.01&                & 0.08 \\
    V1033 Cyg &    0.04&                & 0.12 \\
       KQ Sco &    0.35&  0.52$\pm$ 0.08&      \\
        Y Sct &    0.13&                & 0.23 \\
       BX Sct &    0.13&                & 0.28 \\
       CK Sct &    0.12&                & 0.21 \\
       CM Sct &    0.07&                & 0.15 \\
       CN Sct &    0.28&                & 0.33 \\
       RU Sct &    0.15&  0.14$\pm$ 0.04& 0.11 \\
       TY Sct &    0.25&                & 0.37 \\
       AA Ser &    0.33&                & 0.41 \\
\hline
\end{tabular}
\end{center}
Remark: GET - \citet{Gen14}; LL - \citet{ll11} 
\end{table}

\section{Conclusions and final remarks.}

We derived abundances of 29 elements in 27 Cepheids located mainly 
within 5-7 kpc from Galactic center, with one star SU Sct situated 
in a poorly sampled region of the thin disc between 4 kpc and the 
Galactic center. We found that in the inner region of our Galaxy the 
abundance distributions of oxygen, some $\alpha$-elements and iron 
show a plateau-like structure, or even positive slope. Despite the small 
number of data points within 4 kpc, this finding may have significant impact 
on the Galactic chemical evolution models that include the Galactic bar, and 
its influence on radial gas flows. Future observing programs aiming at 
obtaining additional high-resolution spectra of faint Cepheids located 
within the 0 - 4 kpc zone in the Galaxy disc would be very valuable.
 
\section*{Acknowledgments}
SMA and SAK acknowledge the SCOPES grant No. IZ73Z0-152485 for 
financial support. We acknowledge the queue observing team at 
CFHT and support staff at the VLT and MPGT, and also J.R.D. 
L\'epine for his valuable comments. Authors are thankful to 
the anonymous referee for her/his comments.

\label{lastpage}

\bsp

\end{document}